\documentclass[12pt]{iopart}

\usepackage{amssymb}
\expandafter\let\csname equation*\endcsname\relax 
\expandafter\let\csname endequation*\endcsname\relax
\usepackage{amsmath}
\usepackage{cite} 
\usepackage{bm}
\usepackage{color}
\usepackage[colorlinks,linkcolor=blue,anchorcolor=blue,citecolor=blue]{hyperref}
\usepackage{rotating}
\usepackage{caption}
\usepackage{adjustbox}
\begin{document}

\title[Efficient and secure quantum network coding based on quantum \ldots]{Efficient and secure quantum network coding based on quantum full homomorphic encryption}

\author{Ning Wang$^{1,2}$, Fei Gao$^{3,*}$ \& Song Lin$^{1,*}$}

\address{$^1$ College of Computer and Cyber Security, Fujian Normal University, Fuzhou 350117, China \\
$^2$ School of Information Science and Technology, Zhengzhou Normal University, Zhengzhou, 450044, China\\
$^3$ State Key Laboratory of Networking and Switching Technology, Beijing University of Posts and Telecommunications, Beijing, 100876, China \\
$^*$ Author to whom any correspondence should be addressed.}
\ead{gaof@bupt.edu.cn, and lins95@gmail.com}
\vspace{10pt}
\begin{indented}
\item[]August 2017
\end{indented}

\begin{abstract}
Based on $d$-dimensional quantum full homomorphic encryption, an efficient and secure quantum network coding protocol is proposed in this paper. First, a quantum full homomorphic encryption protocol is constructed utilizing $d$-dimensional universal quantum gates. On this basis, an efficient quantum network coding protocol is proposed. In the protocol, two source nodes encrypt their respective prepared quantum states with the quantum full homomorphic encryption protocol. The two intermediate nodes successively perform homomorphic evaluation of the received quantum states. Finally, the two sink nodes recover the quantum states transmitted by the two source nodes in the butterfly network depending on their measurement results. The performance analysis shows that the proposed quantum network coding protocol is correct and resistant to attacks launched by dishonest intermediate nodes and external eavesdroppers. Compared to related protocols, the proposed protocol not only enables to transfer information in $d$-dimensional quantum system, but also requires only 1 quantum gate and a key of length 2 in the encryption phase, which makes the protocol has higher efficiency.
\end{abstract}

%
%
%
%
%

\section{Introduction}\label{sec1}

with the rapid development of wireless communication technology, we are moving towards the era of 6G wireless networks. Quantum 6G wireless networks \cite{bibWRA,bibRKG} become one of the high-profile research areas, which takes advantage of quantum technology to provide a higher level of data security and reliability for wireless communications. As one of the key technologies for quantum 6G wireless networks, quantum network coding plays an important role.

Network coding, since it was first proposed by Ahlswede et al. \cite{bibAC} in 2000, has made significant advances in theory \cite{bibLY} and applications \cite{bibSS}. It allows intermediate nodes in networks to encode multiple data. In classical networks, intermediate nodes only have storage and forwarding capabilities, and encoding operations only occur at the receiving and sending ends. Therefore, compared to traditional routing \cite{bibSC}, network coding can improve network throughput and save bandwidth. However, with the rapid development of quantum information and quantum computing, the transmission of classical networks has faced new challenges. It has prompted the gradual introduction of classical network coding with powerful throughput and robustness into the quantum field, forming quantum network coding (QNC), which aims to improve the efficiency of quantum communication in quantum networks. Different from classical network coding, the security of quantum network coding is guaranteed by the quantum mechanics principle. It is one of the reasons why quantum network coding has become a hot research topic.

In 2007, Hayashi et al. \cite{bibHI} first introduced the concept of QNC and proposed a QNC protocol based on universal cloning \cite{bibBH}. The protocol implements crossing two qubits, i.e. XQQ protocol, in the butterfly network, but its fidelity is less than 1. To optimise this protocol, Hayashi et al. \cite{bibHM} proposed a QNC scheme with a priori entanglement in the same year. It used two pairs of Bell states shared in advance by the two senders to achieve perfect transmission of two qubits in the butterfly network. Since then, more and more scholars have started to focus on this area, and various QNC protocols \cite{bibKG,bibMC,bibSZ,bibLC,bibLL,
bibSP,bibSL,bibSLL,bibOK,bibCZ,bibCC,bibKO,bibLS,bibSH,bibSHM} have been proposed. In 2009, Kobayashi et al. \cite{bibKG} introduced classical communication into quantum network coding, and achieved lossless quantum state transmission with fidelity of 1 by simulating a classical linear network coding scheme. Subsequently, by sharing non-maximal entangled states among the senders, Ma et al. \cite{bibMC} came up with an efficient QNC protocol for transmitting M-qudit states over the butterfly network. In 2014, Shang et al. \cite{bibSZ} put forward a controlled quantum network coding scheme based on the XQQ protocol and priori entanglement. The scheme made use of the GHZ controlled teleportation.

The above studies \cite{bibHI,bibHM,bibKG,bibMC,bibSZ} did not discuss the security of QNC when attackers attack quantum networks. Since improving the security of QNC is one of the most essential requirements for the development of quantum networks, scholars have started to analyse the security of QNC. In 2016, Shang et al. \cite{bibSP} presented a QNC scheme based on homomorphic signatures. The scheme can resist pollution attacks, it improves the security of quantum network coding. In 2018, Owair et al. \cite{bibOK} considered the possibility of external eavesdroppers attacking quantum networks, and proposed a single-shot secure QNC protocol on a butterfly network with free and open communication. In 2021, Chen et al. \cite{bibCZ} came up with a controlled QNC scheme based on quantum walks. The scheme initially realised the entanglement distribution of butterfly networks, which reduced entanglement resources and enhanced scalability. In addition, the scheme not only analyses external attacks but also discusses possible internal attacks (dishonest intermediate nodes). Recently, Cheng et al. \cite{bibCC} pointed out that the protocol \cite{bibHM} is insecure under the assumption that intermediate nodes are dishonest. They then put forward an improved secure XQQ protocol based on $2$-dimensional quantum homomorphic encryption. Although the protocol can resist both external and internal attacks, it is not efficient enough as the protocol requires 2 quantum gates and a key of length 8 during the homomorphic encryption phase. Furthermore, most of the above protocols are implemented in 2-dimensional quantum systems, which are less universal and information transmission efficient (considering information capacity issues) compared to $d$-dimensional quantum system. Therefore, it is worthwhile to investigate how to design an efficient and secure quantum network coding protocol in a $d$-dimensional quantum system.

Based on the above research on QNC protocols, this paper proposes an efficient and secure QNC protocol based on $d$-dimensional quantum full homomorphic encryption (QFHE). First, we propose the first QFHE protocol with $d$-dimensional universal quantum gates. Further, in accordance with the QFHE protocol, we propose an efficient QNC protocol. In this protocol, firstly, two source nodes encrypt their respective prepared particles using the constructed QFHE protocol during the communication between the two source nodes and the first intermediate node. They send the encrypted particles to the first intermediate node for homomorphic evaluation. Next, the first intermediate node sends the homomorphic evaluated particles to the second intermediate node for homomorphic evaluation. Then, The second intermediate node sends the homomorphic evaluated particles to two sink nodes respectively. Finally, the two sink nodes recover respectively the quantum states transmitted by the two source nodes based on the measurement results, thus enabling perfect transmission of the quantum states over the butterfly network. The protocol analysis shows that the proposed QNC protocol can resist both external and internal attacks. Compared with Cheng et al.'s protocol \cite{bibCC}, the proposed protocol not only achieves the transmission of particles in $d$-dimensional quantum system, but also, and more importantly, requires only 1 quantum gate and a key of length 2 to encode quantum states in the encryption phase, all of which greatly improve the efficiency of the protocol.

The rest of the paper is organized as follows. In Sec. \ref{sec2}, we briefly introduce the relevant preliminaries. Then, a $d$-dimensional quantum full homomorphic encryption protocol is first constructed, furthermore, an effcient and secure quantum network coding protocol is proposed. in Sec. \ref{sec3}. In Sec. \ref{sec4}, the correctness and security of the proposed QNC protocol are analyzed. Finally, a brief discussion and conclusion is given in Sec. \ref{sec5}.

\section{Preliminaries}\label{sec2}

\subsection{$d$-dimensional quantum system}\label{subsec2.1}

In a $d$-dimensional quantum system $H=\mathbb{C}^{d}$, there exists a set of computational basis $B=\{\vert0\rangle,\vert1\rangle,\ldots,\vert d-1\rangle\}$, which is used to measure a single qudit. In addition, there exists a set of Bell basis (EPR pairs) \cite{bibBB}, which is denoted as

\begin{equation}
\vert\psi({{m}_{1}},{{m}_{2}})\rangle =\frac{1}{\sqrt{d}}\sum\limits_{j=0}^{d-1}\omega^{jm_{1}}\vert j,j+{{m}_{2}}\rangle,
\label{eq:1}
\end{equation}
where, $m_{1},m_{2}\in \mathbb{Z}_{d}$, $\omega=e^{2\pi i/d}$, ``+'' denotes mod $d$ add. Quantum measurement in the Bell basis is called Bell measurement. When $m_{1}=m_{2}=0$, $\vert\psi_{00}\rangle =\frac{1}{\sqrt{d}}\sum\limits_{j=0}^{d-1}\vert j,j\rangle$, which is used as the quantum entanglement resource for the quantum network coding protocol proposed in Sec. \ref{subsec3.2}. Bell measurement is commonly used to jointly measure two-particle state over $H=\mathbb{C}^{d}$. It can be used to achieve quantum teleportation \cite{bibBB}, which is described below.

\begin{figure}[h]%
\centering
\includegraphics[width=0.7\textwidth]{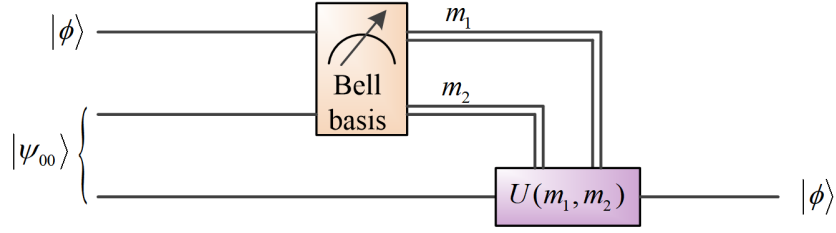}
\caption{The quantum teleportation circuit of a qudit. The double line indicates transmission of classical dits.}\label{fig:1}
\end{figure}

Assume that the sender, Alice has a $d$-dimensional single-particle state
\begin{equation}
\left| \phi  \right\rangle =\sum\limits_{j=0}^{d-1}{{{\alpha }_{j}}\left| j \right\rangle },
\label{eq:2}
\end{equation}
and that she and Bob each possess one of the particles of the Bell state $\vert\psi_{00}\rangle$. The state of the whole system can be written as

\begin{equation}
\left| \phi  \right\rangle \left| {{\psi }_{00}} \right\rangle =\frac{1}{\sqrt{d}}\sum\limits_{j,k=0}^{d-1}{{{\alpha }_{j}}{{\left| j \right\rangle }_{1}}} {{{\left| k \right\rangle }_{2}}{{\left| k \right\rangle }_{3}}},
\label{eq:3}
\end{equation}
where, Alice has particles 1 and 2, Bob has particle 3. If Alice performs the Bell measurement on the particles 1 and 2, she obtains the measurements $m_1$ and $m_2$. Simultaneously, the particle 3 owned by Bob collapses to

\begin{equation}
{{U}^{\dagger}}({{m}_{1}},{{m}_{2}})\left| \phi  \right\rangle_3 =\frac{1}{\sqrt{d}}\sum\limits_{j=0}^{d-1}{{{\alpha }_{j}}{{\omega }^{-j{{m}_{1}}}}}{{\left| j+{{m}_{2}} \right\rangle }_{3}}.
\label{eq:4}
\end{equation}
Here, ${{U}^{\dagger}}({{m}_{1}},{{m}_{2}})$ is the conjugate transpose of the unitary operation

\begin{equation}
U({{m}_{1}},{{m}_{2}})=\sum\limits_{j=0}^{d-1}{{{\omega }^{j{{m}_{1}}}}}\left| j \right\rangle \left\langle  j+{{m}_{2}} \right|.
\label{eq:5}
\end{equation}
Once Bob learns Alice's measurements $m_1$ and $m_2$, he can recover the quantum state $\vert \phi\rangle$ transmitted by Alice by performing the unitary operation $U({{m}_{1}},{{m}_{2}})$ on the particle 3. The quantum teleportation circuit between Alice and Bob is shown in Figure \ref{fig:1}.

Next, we will describe some qudit gates involved in this paper. In $H=\mathbb{C}^{d}$, the Clifford gate includes the single-qudit Pauli operations, Hadamard gate $H$, the phase gate $S$, and the two-qudit controlled-$X$ gate $CX$. The single-qudit Pauli operations consist of $X$, $Y$ and $Z$ gates, which are defined \cite{bibMJ} as follows:

\begin{equation}
\begin{aligned}
& X=\sum\limits_{j=0}^{d-1}{\left| j+1 \right\rangle \left\langle  j \right|}, \\
& Y=-\sum\limits_{j=0}^{d-1}\omega^{(2j-1)/2}{\left| j+1 \right\rangle \left\langle  j \right|}, \\
& Z=\sum\limits_{j=0}^{d-1}{{{\omega }^{j}}}\left| j \right\rangle \left\langle  j \right|.
\end{aligned}
\label{eq:6}
\end{equation}
$X$, $Y$ and $Z$ satisfy the properties: ${{X}^{d}}={{Y}^{d}}={{Z}^{d}}={{I}_{d}}$, $Y=-{{\omega }^{-1/2}}XZ$. The Hadamard gate $H$ and the phase gate $S$ are denoted \cite{bibMJ} as

\begin{equation}
\begin{aligned}
& H=\frac{1}{\sqrt{d}}\sum\limits_{j=0}^{d-1}{\sum\limits_{k=0}^{d-1}{{{\omega }^{jk}}\left| k \right\rangle \left\langle  j \right|,}} \\
 & S=\sum\limits_{j=0}^{d-1}{{{\omega }^{{(j-d+2)j}/{2}\;}}\left| j \right\rangle \left\langle  j \right|}.
\end{aligned}
\label{eq:7}
\end{equation}
The two-qudit controlled-$X$ gate is described \cite{bibMJ} as follows:

\begin{equation}
\begin{aligned}
&CX=\sum\limits_{j=0}^{d-1}{\left| j \right\rangle \left\langle  j \right|\otimes {{X}^{j}}}.
\end{aligned}
\label{eq:8}
\end{equation}
Here, the operation $X^{j}$ on the second qudit of $CX$ is controlled by the value $j$ of the first qudit. In addition to Clifford gates, this section also introduces a non-Clifford gate, i.e. qudit $\pi/8$ gate $T_{t}$. When $d>3$, $T_{t}$ can be expressed \cite{bibHV} as

\begin{equation}
{{T}_{t}}=T({{t}_{0}},{{t}_{1}},\ldots )=\sum\limits_{j=0}^{d-1}{{{\omega }^{{{t}_{j}}}}\left| j \right\rangle \left\langle  j \right|({{t}_{j}}\in {{\mathbb{Z}}_{d}})}.
\label{eq:9}
\end{equation}

The above given quantum gates are collectively known as universal quantum gates, and they can be used to construct any universal quantum circuit. Without considering the global phase, there are the following conditional commutation relations among $X,\ Y,\ Z$, $H$, $S$, $T_{t}$, and $CX$,

\begin{equation}
\left\{ \begin{aligned}
& X{{X}^{p}}{{Z}^{q}}={{X}^{p}}{{Z}^{q}}X, \\
 & Y{{X}^{p}}{{Z}^{q}}={{X}^{p}}{{Z}^{q}}Y, \\
 & Z{{X}^{p}}{{Z}^{q}}={{X}^{p}}{{Z}^{q}}Z, \\
 & H{{X}^{p}}{{Z}^{q}}={{X}^{q}}{{Z}^{p}}H, \\
 & S{{X}^{p}}{{Z}^{q}}={{X}^{q}}{{Z}^{p+q}}S, \\
 & {{X}^{r}}{{Z}^{r^\prime }}{{S}^{p}}{{T}_{t}}{{X}^{p}}{{Z}^{q}}={{X}^{p+r}}{{Z}^{p+q+r^\prime}}T_{t}, \\
 & C{{X}_{1,2}}\left( {{X}^{p}}{{Z}^{q}}\otimes {{X}^{s}}{{Z}^{t}} \right)=\left( {{X}^{p}}{{Z}^{q+t}}\otimes {{X}^{p+s}}{{Z}^{t}} \right)C{{X}_{1,2}}. \\
\end{aligned} \right.
\label{eq:10}
\end{equation}
Here, $p,q,s,t,r,r^\prime\in {{\mathbb{Z}}_{d}}$. Eq. (\ref{eq:10}) is the basis for the QFHE protocol presented in Sec. \ref{subsec3.1}. Furthermore, since there is a constant relationship between X, Z and $U({{m}_{1}},{{m}_{2}})$ as follows:

\begin{equation}
U({{m}_{1}},{{m}_{2}})={{\left( {{X}^{\dagger}} \right)}^{{{m}_{2}}}}{{Z}^{{{m}_{1}}}}.
\label{eq:11}
\end{equation}
Therefore, in order to reduce the communication complexity, the unitary operation $U$ is used as the homomorphic encryption algorithm in proposed QNC protocol in Sec. \ref{subsec3.2}.

\subsection{The XQQ protocol based on quantum teleportation} \label{subsec2.2}

\begin{figure}[h]%
\centering
\includegraphics[width=5.5in]{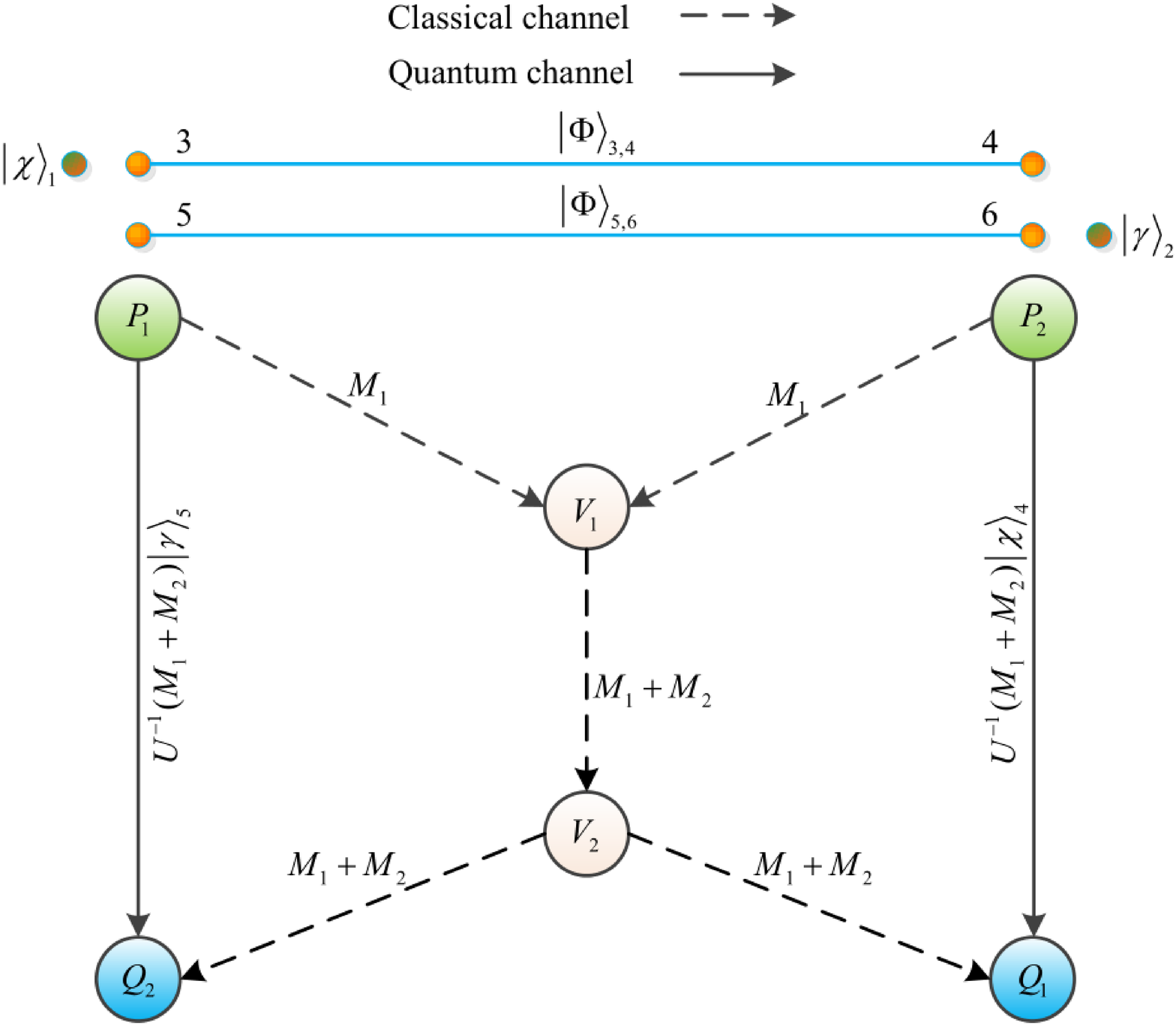}
\caption{The butterfly network of Hayashi's protocol.}\label{fig:2}
\end{figure}

In this section, we briefly introduce the XQQ protocol proposed by Hayashi \cite{bibHM}. In this protocol, there are two source nodes $P_1$ and $P_2$, two intermediate nodes $V_1$ and $V_2$, and two sink nodes $Q_1$ and $Q_2$. It is assumed that $P_1$ and $P_2$ pre-share two pairs of maximally entangled states ${{\left| \Phi  \right\rangle }_{3,4}}$ and ${{\left| \Phi  \right\rangle }_{5,6}}$, where $P_1$ has particles 3 and 5, $P_2$ has particles 4 and 6. The quantum states transmitted by $P_1$ and $P_2$ are ${{\left| \chi  \right\rangle }_{1}}$ and ${{\left| \chi  \right\rangle }_{2}}$ respectively. The butterfly network of the protocol is shown in Figure \ref{fig:2}.

In Hayashi's \cite{bibHM} protocol, the two intermediate nodes $V_1$ and $V_2$ can be considered as internal participants since they can perform encoding operations. $V_1$ and $V_2$ are by default honest and they are not aware of the encryption rules between the two source nodes and the two sink nodes. However, in the practical application, the intermediate nodes may be dishonest and he tries to launch attack to recover out the quantum states sent by the source nodes. In addition, the intermediate nodes may also be the source nodes of another butterfly network in a certain large quantum network, hence the consistency of the encryption rules will make the network more universal.

\begin{figure}[htp]%
\centering
\includegraphics[width=5.5in]{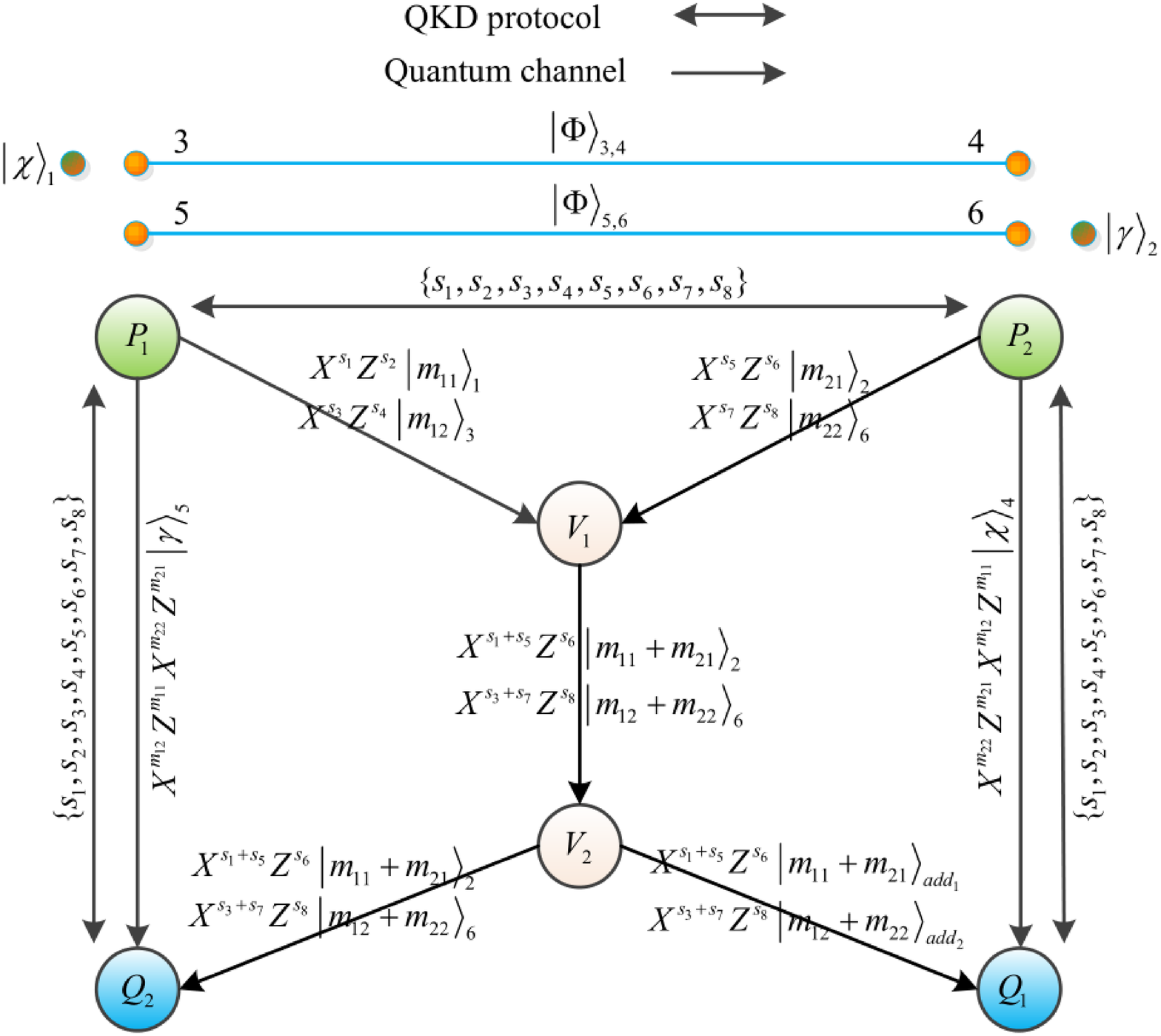}
\caption{The butterfly network of Cheng et al.'s protocol.}\label{fig:3}
\end{figure}

Based on the above two situations, Cheng et al. \cite{bibCC} pointed out that Hayashi's protocol \cite{bibHM} is insecure, and proposed an improved XQQ protocol. The butterfly network of this protocol is shown in Figure \ref{fig:3}.
The protocol exploits 2-dimensional homomorphic encryption scheme to encode measurement particles and achieve secure transmission of measurement results, thus resisting attacks by dishonest intermediate nodes. However, the protocol requires 2 quantum gates and a key of length 8 in the encryption phase, which  makes the protocol less efficient. In addition, compared to $d$-dimensional quantum systems, the protocol is implemented using 2-dimensional quantum systems, which are not sufficiently universal and information transfer efficient. In response to these problems, we propose an efficient and secure QNC protocol based on $d$-dimensional QFHE in next section.

\section{Quantum network coding protocol based on quantum full homomorphic encryption} \label{sec3}

Considering that the existing $d$-dimensional QHE protocol \cite{bibZS} only involves single-particle quantum gate operations and is not applicable to arbitrary quantum gates. It cannot satisfy the needs of the QNC protocol proposed in Sec. \ref{subsec3.2}. Therefore, this section first constructs a QFHE protocol with $d$-dimensional universal quantum gates. Then, based on the proposed QFHE protocol, we propose an efficient and secure quantum network coding protocol. The correctness of both schemes has been theoretically proven.

\subsection{Quantum full homomorphic encryption protocol} \label{subsec3.1}

In this section, a QFHE protocol with $d$-dimensional universal quantum gates is constructed, which is a a four-tuple of quantum algorithms (key generation, encryption, evaluation, decryption) \cite{bibLM,bibZS}. The client generates encryption keys and decryption keys using the key generation algorithm. He uses the encryption algorithm to encrypt quantum state plaintext to generate quantum state ciphertext. After that, the client sends the quantum state ciphertext to the server via the quantum-secure channel. After receiving the quantum state ciphertext, the server executes the evaluation algorithm. He then sends the calculation result to the client. The client decrypts the calculation result sent back by the server, and it is the calculation result expected by the client. The set of quantum operations allowed by the protocol is $F=\{X,Y,Z,H,S, T_{t},CX\}$. The proposed QFHE protocol is described in detail as follows.

(1) Key generation ($KeyGen_{\Delta}$): When a single (double) qudit gate $G\in\{X,Y,Z,H,S,T_{t}\}$($CX$) acts on the quantum state plaintext $\sigma_{1}$ ($\sigma_{1}$ and $\sigma_{2}$), the client randomly generates the encryption keys $p, q\in\mathbb{Z}_{d}$ ($p,\ q,\ s,\ t\in\mathbb{Z}_{d}$) according to the key generation algorithm.

(2) Encryption ($Enc_{\Delta}$): Based on the keys $p$, $q$ ($p$, $q$, $s$, $t$), the encryption algorithm $Enc_{\Delta}$ performs encryption on the quantum state plaintext $\sigma_{1}$ ($\sigma_{1}$, $\sigma_{2}$). The obtained quantum state ciphertext is (are) $\rho_{c_{1}}=X^{p}Z^{q}\sigma_{1}(X^{p}Z^{q})^{\dagger}$ ($\rho_{c_{1}}=X^{p}Z^{q}\sigma_{1}(X^{p}Z^{q})^{\dagger}$ and $\rho_{c_{2}}=X^{p}Z^{q}\sigma_{2}(X^{p}Z^{q})^{\dagger}$). Finally, the client sends the quantum state ciphertext $\rho_{c_{1}}$ ($\rho_{c_{1}}$ and $\rho_{c_{2}}$) to the server via a secure quantum channel.

(3) Evaluate ($Eval_{\Delta}$): After receiving the quantum state ciphertext $\rho_{c_{1}}$ ($\rho_{c_{1}}$ and $\rho_{c_{2}}$) from the client, the server performs the evaluation operation on it (them). The acquired evaluation result is $\rho'=Eval_{\Delta}(G\in\{X,Y,Z,H,S,T_{t}\},\rho_{c_{1}})=G\rho_{c_{1}}G^{\dagger}$
($\rho''=Eval_{\Delta}(CX_{1,2},\rho_{c_{1}},\rho_{c_{2}})=CX_{1,2}(\rho_{c_{1}}\otimes
\rho_{c_{2}})CX_{1,2}^{\dagger}$). He sends $\rho'$ ($\rho''$) to the client via a secure quantum channel.

(4) Decryption ($Dec_{\Delta}$): Upon receiving the evaluation result $\rho'$ ($\rho''$) from the server, the client performs the decryption operation on it with the decryption keys $p', q'$ ($p',\ q',\ s',\ t'$). That is, $\sigma'=Dec_{\Delta}(p',q',\rho')=Z^{-q'}X^{-p'}\rho'
(Z^{-q'}X^{-p'})^{\dagger}$ ($\sigma''=Dec_{\Delta}(p',q',s',t',\rho'')=(Z^{-q'}X^{-p'}\otimes Z^{-t'}X^{-s'})\rho''
(Z^{-q'}X^{-p'}\otimes Z^{-t'}X^{-s'})^{\dagger}$). The decryption keys is generated by the key generation algorithm.

The scheme takes advantage of the conditional interchangeability between qudit gates, and exchanges the evaluation operator and the encryption operator by means of equivalent multiplication to complete homomorphic encryption. The correctness of the protocol is guaranteed by the calculation of the evaluation and decryption results. The specific correctness analysis is shown below.

\textbf{Theorem 1} When the evaluation operators $X,\ Y,\ Z,\ H,\ S,\ T_{t}$ and $CX$ act respectively on the quantum state ciphertext, the following equations are obtained.

\begin{equation}
\begin{aligned}
 \rho^{\prime }_{1}& =Eva{{l}_{\Delta }}\left( X,{{\rho }_{{{c}_{1}}}} \right)={{X}^{p}}{{Z}^{q}}X{{\sigma }_{1}}{{X}^{\dagger }}{{({{X}^{p}}{{Z}^{q}})}^{\dagger }}, \\
 \rho^{\prime }_{2}&=Eva{{l}_{\Delta }}\left( Y,{{\rho }_{{{c}_{1}}}} \right)={{X}^{p}}{{Z}^{q}}Y{{\sigma }_{1}}{{Y}^{\dagger }}{{({{X}^{p}}{{Z}^{q}})}^{\dagger }}, \\
 \rho^{\prime }_{3}&=Eva{{l}_{\Delta }}\left( Z,{{\rho }_{{{c}_{1}}}} \right)={{X}^{p}}{{Z}^{q}}Z{{\sigma }_{1}}{{Z}^{\dagger }}{{({{X}^{p}}{{Z}^{q}})}^{\dagger }}, \\
 \rho^{\prime }_{4}&=Eva{{l}_{\Delta }}\left( H,{{\rho }_{{{c}_{1}}}} \right)={{X}^{q}}{{Z}^{p}}H{{\sigma }_{1}}{{H}^{\dagger }}{{({{X}^{q}}{{Z}^{p}})}^{\dagger }}, \\
 \rho^{\prime }_{5}&=Eva{{l}_{\Delta }}\left( S,{{\rho }_{{{c}_{1}}}} \right)={{X}^{p}}{{Z}^{q}}Z{{\sigma }_{1}}{{Z}^{\dagger }}{{({{X}^{p}}{{Z}^{q}})}^{\dagger }}, \\
 \rho^{\prime }_{6}&=Eva{{l}_{\Delta }}\left( {{T}_{t}},{{\rho }_{{{c}_{1}}}} \right)={{X}^{q}}{{Z}^{p}}{{T}_{t}}{{\sigma }_{1}}{{T}_{t}}^{\dagger }{{({{X}^{q}}{{Z}^{p}})}^{\dagger }}, \\
 \rho^{\prime \prime} &=Eva{{l}_{\Delta }}\left( C{{X}_{1,2}},{{\rho }_{{{c}_{1}}}},{{\rho }_{{{c}_{2}}}} \right) \\
 & =\left( {{X}^{p}}{{Z}^{q+t}}\otimes {{X}^{p+s}}{{Z}^{t}} \right)C{{X}_{1,2}}\left( {{\sigma }_{1}}\otimes {{\sigma }_{2}} \right)CX_{1,2}^{\dagger }{{\left( {{X}^{p}}{{Z}^{q+t}}\otimes {{X}^{p+s}}{{Z}^{t}} \right)}^{\dagger }}.
\label{eq:12}
\end{aligned}
\end{equation}

\textbf{proof} Based on the conditional exchangeability between quantum gates in Eq. (\ref{eq:10}), we get

\begin{equation}
\begin{aligned}
 \rho {^{\prime }_{1}}&=Eva{{l}_{\Delta }}\left( X,{{\rho }_{{{c}_{1}}}} \right)=X{{\rho }_{{{c}_{1}}}}{{X}^{\dagger }}=X{{X}^{p}}{{Z}^{q}}{{\sigma }_{1}}{{({{X}^{p}}{{Z}^{q}})}^{\dagger }}{{X}^{\dagger }} \\
 & =X{{X}^{p}}{{Z}^{q}}{{\sigma }_{1}}{{(X{{X}^{p}}{{Z}^{q}})}^{\dagger }}={{X}^{p}}{{Z}^{q}}X{{\sigma }_{1}}{{X}^{\dagger }}{{({{X}^{p}}{{Z}^{q}})}^{\dagger }}, \\
 \rho {^{\prime }_{2}}&=Eva{{l}_{\Delta }}\left( Y,{{\rho }_{{{c}_{1}}}} \right)=Y{{\rho }_{{{c}_{1}}}}{{Y}^{\dagger }}=Y{{X}^{p}}{{Z}^{q}}{{\sigma }_{1}}{{({{X}^{p}}{{Z}^{q}})}^{\dagger }}{{Y}^{\dagger }} \\
 & ={{X}^{p}}{{Z}^{q}}Y{{\sigma }_{1}}{{Y}^{\dagger }}{{({{X}^{p}}{{Z}^{q}})}^{\dagger }}, \\
 \rho {^{\prime }_{3}}&=Eva{{l}_{\Delta }}\left( Z,{{\rho }_{{{c}_{1}}}} \right)=Z{{\rho }_{{{c}_{1}}}}{{Z}^{\dagger }}=Z{{X}^{p}}{{Z}^{q}}{{\sigma }_{1}}{{({{X}^{p}}{{Z}^{q}})}^{\dagger }}{{Z}^{\dagger }} \\
 & ={{X}^{p}}{{Z}^{q}}Z{{\sigma }_{1}}{{Z}^{\dagger }}{{({{X}^{p}}{{Z}^{q}})}^{\dagger }}, \\
 \rho {^{\prime }_{4}}&=Eva{{l}_{\Delta }}\left( H,{{\rho }_{{{c}_{1}}}} \right)=H{{\rho }_{{{c}_{1}}}}{{H}^{\dagger }}=H{{X}^{p}}{{Z}^{q}}{{\sigma }_{1}}{{({{X}^{p}}{{Z}^{q}})}^{\dagger }}{{H}^{\dagger }} \\
 & ={{X}^{q}}{{Z}^{p}}H{{\sigma }_{1}}{{H}^{\dagger }}{{({{X}^{q}}{{Z}^{p}})}^{\dagger }}, \\
 \rho {^{\prime }_{5}}&=Eva{{l}_{\Delta }}\left( S,{{\rho }_{{{c}_{1}}}} \right)=S{{\rho }_{{{c}_{1}}}}{{S}^{\dagger }}=S{{X}^{p}}{{Z}^{q}}{{\sigma }_{1}}{{({{X}^{p}}{{Z}^{q}})}^{\dagger }}{{S}^{\dagger }} \\
 & ={{X}^{p}}{{Z}^{p+q}}S{{\sigma }_{1}}{{S}^{\dagger }}{{({{X}^{p}}{{Z}^{p+q}})}^{\dagger }}, \\
 \rho {^{\prime }_{6}}&=Eva{{l}_{\Delta }}\left( {{T}_{t}},{{\rho }_{{{c}_{1}}}} \right)={{X}^{r}}{{Z}^{r^\prime }}{{S}^{p}}{{T}_{t}}{{\rho }_{{{c}_{1}}}}{{({{X}^{r}}{{Z}^{r^\prime }}{{S}^{p}}{{T}_{t}})}^{\dagger }} \\
 & ={{X}^{r}}{{Z}^{r^\prime }}{{S}^{p}}{{T}_{t}}{{X}^{p}}{{Z}^{q}}{{\sigma }_{1}}{{({{X}^{p}}{{Z}^{q}})}^{\dagger }}{{({{X}^{r}}{{Z}^{r^\prime }}{{S}^{p}}{{T}_{t}})}^{\dagger }} \\
 & ={{X}^{p+r}}{{Z}^{p+q+r^\prime }}{{T}_{t}}{{\sigma }_{1}}{{T}_{t}}^{\dagger }{{({{X}^{p+r}}{{Z}^{p+q+r^\prime }})}^{\dagger }},
\end{aligned}
\label{eq:13-1}
\end{equation}

\begin{equation}
\begin{aligned}
 \rho^{\prime \prime}& =Eva{{l}_{\Delta }}\left( C{{X}_{1,2}},{{\rho }_{{{c}_{1}}}},{{\rho }_{{{c}_{2}}}} \right)=C{{X}_{1,2}}\left( {{\rho }_{{{c}_{1}}}}\otimes {{\rho }_{{{c}_{2}}}} \right)CX_{1,2}^{\dagger} \\
 & =C{{X}_{1,2}}\left( U(p,q){{\sigma }_{1}}{{U}^{\dagger }}(p,q)\otimes U(s,t){{\sigma }_{2}}{{U}^{\dagger }}(s,t) \right)CX_{1,2}^{\dagger } \\
 & =C{{X}_{1,2}}\left( U(p,q)\otimes U(s,t) \right)\left( {{\sigma }_{1}}\otimes {{\sigma }_{2}} \right){{\left( U(p,q)\otimes U(s,t) \right)}^{\dagger }}CX_{1,2}^{\dagger } \\
 & =\left( U(p+s,q)\otimes U(s,q+t) \right)C{{X}_{1,2}}\left( {{\sigma }_{1}}\otimes {{\sigma }_{2}} \right) \\
 & CX_{1,2}^{\dagger }{{\left( U(p+s,q)\otimes U(s,q+t) \right)}^{\dagger }}.
\end{aligned}
\label{eq:13-2}
\end{equation}
From Eq. (\ref{eq:13-1}) and (\ref{eq:13-2}), it is clear that when the evaluation operator acts on the quantum state ciphertext, it is equivalent to directly encrypting the evaluation operator on the quantum state plaintext. Therefore the evaluation algorithm for the proposed QFHE protocol is correct.

\textbf{Theorem 2} The decryption algorithm $De{{c}_{\Delta }}$ is correct when the following equation holds.

\begin{equation}
\begin{aligned}
 & \sigma^{\prime} =De{{c}_{\Delta }}\left( p^{\prime},q^{\prime},\rho^{\prime} \right)=G{{\sigma }_{1}}{{G}^{\dagger}}, \\
 & \sigma^{\prime\prime} =De{{c}_{\Delta }}\left( p^{\prime},q^{\prime} ,s^{\prime},t^{\prime},\rho^{\prime\prime}  \right)=C{{X}_{1,2}}\left( {{\sigma }_{1}}\otimes {{\sigma }_{2}} \right)CX_{1,2}^{\dagger}. \\
\end{aligned}
\label{eq:14}
\end{equation}

\textbf{Proof} According to Theorem 1 and the conditional interchangeability between qudit gates in Eq. (\ref{eq:10}), we obtain the homomorphic transformation results of the evaluation algorithm. When the decryption algorithm $Dec_{\bigtriangleup}$ decrypts the evaluation result of the quantum state ciphertext, we discuss it in the following several cases.

(1) When the evaluation operator $G\in\{X, Y, Z\}$, the decryption key is the same as the encryption keys based on Eq. (\ref{eq:10}), i.e. $\left( p^{\prime},q^{\prime} \right)=\left( p,q \right)$. Then the decryption results corresponding to the evaluation operators $X$, $Y$ and $Z$ are

\begin{equation}
\begin{aligned}
 \sigma {^{\prime}_{1}}&=De{{c}_{\Delta }}\left( p^\prime ,q^\prime ,\rho {{^\prime }_{1}} \right)={{Z}^{-q}}{{X}^{-p}}\rho {{^\prime }_{1}}{{({{Z}^{-q}}{{X}^{-p}})}^{\dagger }} \\
 & ={{Z}^{-q}}{{X}^{-p}}{{X}^{p}}{{Z}^{q}}X{{\sigma }_{1}}{{X}^{\dagger }}{{({{X}^{p}}{{Z}^{q}})}^{\dagger }}{{({{Z}^{-q}}{{X}^{-p}})}^{\dagger }} \\
 & =X{{\sigma }_{1}}{{X}^{\dagger }}, \\
 \sigma {{^\prime }_{2}}&=De{{c}_{\Delta }}\left( p^\prime ,q^\prime ,\rho {{^\prime }_{2}} \right)={{Z}^{-q}}{{X}^{-p}}\rho {^{\prime }_{2}}{{({{Z}^{-q}}{{X}^{-p}})}^{\dagger }} \\
 & ={{Z}^{-q}}{{X}^{-p}}{{X}^{p}}{{Z}^{q}}Y{{\sigma }_{1}}{{Y}^{\dagger }}{{({{X}^{p}}{{Z}^{q}})}^{\dagger }}{{({{Z}^{-q}}{{X}^{-p}})}^{\dagger }} \\
 & =Y{{\sigma }_{1}}{{Y}^{\dagger }}. \\
 \sigma {^{\prime }_{3}}&=De{{c}_{\Delta }}\left( p^\prime ,q^\prime ,\rho {^{\prime }_{3}} \right)={{Z}^{-q}}{{X}^{-p}}\rho {^{\prime }_{3}}{{({{Z}^{-q}}{{X}^{-p}})}^{\dagger }} \\
 & ={{Z}^{-q}}{{X}^{-p}}{{X}^{p}}{{Z}^{q}}Z{{\sigma }_{1}}{{Z}^{\dagger }}{{({{X}^{p}}{{Z}^{q}})}^{\dagger }}{{({{Z}^{-q}}{{X}^{-p}})}^{\dagger }} \\
 & =Z{{\sigma }_{1}}{{Z}^{\dagger }}.
\end{aligned}
\label{eq:15-1}
\end{equation}

(2) When the evaluation operator $G=H$, the decryption key is updated as $\left( p^{\prime},q^{\prime} \right)=\left(q, p \right)$ according to Eq. (\ref{eq:10}). Then the decryption result

\begin{equation}
\begin{aligned}
 \sigma {^{\prime }_{4}}&=De{{c}_{\Delta }}\left( p^\prime ,q^\prime ,\rho {^{\prime }_{4}} \right)={{Z}^{-p}}{{X}^{-q}}\rho {^{\prime }_{4}}{{({{Z}^{-p}}{{X}^{-q}})}^{\dagger }} \\
 & ={{Z}^{-p}}{{X}^{-q}}{{X}^{q}}{{Z}^{p}}H{{\sigma }_{1}}{{H}^{\dagger }}{{({{X}^{q}}{{Z}^{p}})}^{\dagger }}{{({{Z}^{-p}}{{X}^{-q}})}^{\dagger }} \\
 & =H{{\sigma }_{1}}{{H}^{\dagger }}.
\end{aligned}
\label{eq:15-2}
\end{equation}

(3) When the evaluation operator $G=S$, the decryption key is renewed as $\left( p^{\prime},q^{\prime} \right)=\left( p, p+q \right)$ depending on Eq. (\ref{eq:10}). Then the decryption result is

\begin{equation}
\begin{aligned}
 \sigma {^{\prime }_{5}}&=De{{c}_{\Delta }}\left( p^\prime ,q^\prime ,\rho {^{\prime }_{5}} \right)={{Z}^{-(p+q)}}{{X}^{-p}}\rho {^{\prime }_{5}}{{\left( {{Z}^{-(p+q)}}{{X}^{-p}} \right)}^{\dagger }} \\
 & ={{Z}^{-\left( p+q \right)}}{{X}^{-p}}{{X}^{p}}{{Z}^{p+q}}S{{\sigma }_{1}}{{S}^{\dagger }}{{\left( {{X}^{p}}{{Z}^{p+q}} \right)}^{\dagger }}{{\left( {{Z}^{-(p+q)}}{{X}^{-p}} \right)}^{\dagger }} \\
 & =S{{\sigma }_{1}}{{S}^{\dagger }}.
\end{aligned}
\label{eq:15-3}
\end{equation}

(4) When the evaluation operator $G=T_{t}$, the decryption key is renewed as $\left( p^{\prime},q^{\prime} \right)=\left( p+r,p+q+r^\prime \right)$ in accordance with Eq. (\ref{eq:10}). Then the decryption result is

\begin{equation}
\begin{aligned}
  \sigma {^{\prime }_{6}}&=De{{e}_{\Delta }}\left( p^\prime ,q^\prime ,\rho {^{\prime }_{6}} \right)={{Z}^{-(p+q+r^\prime )}}{{X}^{-(p+r)}}\rho {^{\prime }_{6}}{{({{Z}^{-(p+q+r^\prime )}}{{X}^{-(p+r)}})}^{\dagger }} \\
 & ={{Z}^{-\left( p+q+r^\prime  \right)}}{{X}^{-\left( p+r \right)}}{{X}^{p+r}}{{Z}^{p+q+r^\prime }}{{T}_{t}}{{\sigma }_{1}}{{T}_{t}}^{\dagger }{{( {{X}^{p+r}}{{Z}^{p+q+r^\prime }} )}^{\dagger }}{{( {{Z}^{-(p+q+r^\prime )}}{{X}^{-(p+r)}} )}^{\dagger }} \\
 & ={{T}_{t}}{{\sigma }_{1}}{{T}_{t}}^{\dagger }.
\end{aligned}
\label{eq:15-4}
\end{equation}

(5) When the evaluation operator is $CX_{1,2}$, the decryption keys are updated to $\left( p^\prime, q^\prime  \right)=\left( p+s,q \right)$ and $\left( s^\prime, t^\prime \right)=\left( s, q+t \right)$ depending on Eq. (\ref{eq:10}). The decryption result is

\begin{equation}
\begin{aligned}
 \sigma^{\prime \prime} &=De{{c}_{\Delta }}\left( p^\prime ,q^\prime ,s^\prime ,t^\prime ,\rho^{\prime \prime}  \right)=\left( {{U}^{\dagger }}\left( p^\prime ,q^\prime  \right)\otimes {{U}^{\dagger }}\left( s^\prime ,t^\prime  \right) \right)\rho^{\prime \prime}
 \left( U\left( p^\prime ,q^\prime  \right)\otimes U\left( s^\prime ,t^\prime  \right) \right) \\
 & =\left( {{U}^{\dagger }}\left( p+s,q \right)\otimes {{U}^{\dagger }}\left( s,q+t \right) \right)\rho^{\prime \prime} \left( U\left( p+s,q \right)\otimes U\left( s,q+t \right) \right) \\
 & =\left( {{U}^{\dagger }}\left( p+s,q \right)\otimes {{U}^{\dagger }}\left( s,q+t \right) \right)\left( U(p+s,q)\otimes U(s,q+t) \right)C{{X}_{1,2}}\left( {{\sigma }_{1}}\otimes {{\sigma }_{2}} \right) \\
 & CX_{1,2}^{\dagger }{{\left( U(p+s,q)\otimes U(s,q+t) \right)}^{\dagger }}\left( U\left( p+s,q \right)\otimes U\left( s,q+t \right) \right) \\
 & =\left( {{U}^{\dagger }}\left( p+s,q \right)U(p+s,q)\otimes {{U}^{\dagger }}\left( s,q+t \right)U(s,q+t) \right)C{{X}_{1,2}}\left( {{\sigma }_{1}}\otimes {{\sigma }_{2}} \right) \\
 & CX_{1,2}^{\dagger }\left( {{U}^{\dagger }}(p+s,q)U\left( p+s,q \right)\otimes {{U}^{\dagger }}(s,q+t)U\left( s,q+t \right) \right) \\
 & =C{{X}_{1,2}}\left( {{\sigma }_{1}}\otimes {{\sigma }_{2}} \right)CX_{1,2}^{+}.
\end{aligned}
\label{eq:16}
\end{equation}

From the above analysis, it can be seen that the direct evaluation result of the quantum state plaintext can be obtained by the decryption algorithm. Therefore, the decryption algorithm of the proposed QFHE protocol is correct.

In summary, according to the above proofs of Theorem 1 and Theorem 2, we fully demonstrate that the proposed QFHE protocol is correct.

\subsection{Quantum network coding protocol} \label{subsec3.2}

\begin{figure}[h]%
\centering
\includegraphics[width=5.5in]{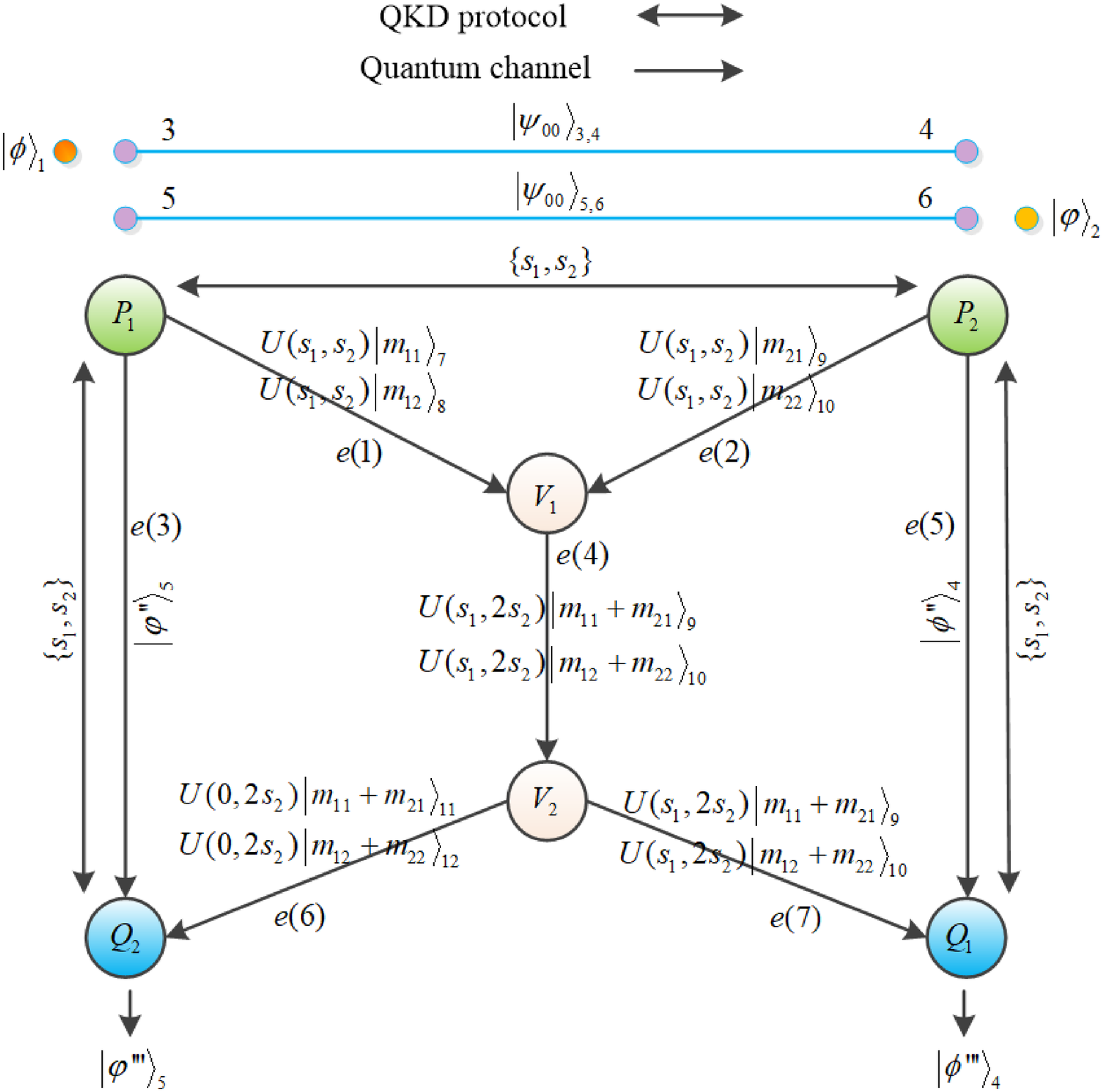}
\caption{The process of quantum secure multi-party summation protocol }\label{fig:4}
\end{figure}

In this section, an efficient and secure QNC protocol is proposed using the constructed $d$-dimensional QFHE protocol, as shown in Figure \ref{fig:4}. The task of this protocol is to transmit the quantum states $\vert\phi\rangle_1$ and $\vert\varphi\rangle_2$ possessed by the two source nodes $P_1$ and $P_2$ to the corresponding sink nodes $Q_1$ and $Q_2$, respectively, through a quantum butterfly network consisting of intermediate nodes $V_1$ and $V_2$ and edges e(1), e(2),...,e(7). In the butterfly network, assume that the two source nodes $P_1$ and $P_2$ as well as the two sink nodes $Q_1$ and $Q_2$ are both honest, they share two common random keys. While the two intermediate nodes $V_1$ and $V_2$ are dishonest, who try to obtain the secret information of the others to recover the transmission quantum states. The specific steps of the protocol are as follows.

\textbf{Step 1: Quantum key distribution process.} $P_1$ and $P_2$ shares two random keys $s_1,\ s_2\in\mathbb{Z}_{d}$ utilising a secure quantum key distribution protocol \cite{bibYZ}. Similarly, the protocol \cite{bibYZ} is executed respectively between $P_1$ and $Q_2$ together with $P_2$ and $Q_1$, such that both $Q_1$ and $Q_2$ obtain information about the keys $s_1,\ s_2$.

\textbf{Step 2: Quantum information transmission process.} The quantum states transmitted by $P_1$ and P2 are $\vert\phi\rangle_1=\sum_{j=0}^{d-1}\alpha_{j}\vert j\rangle_1$ and $\vert\varphi\rangle_2=\sum_{j=0}^{d-1}\beta_{j}\vert j\rangle_2$, respectively. The two pairs of Bell states that they share are $\vert\psi_{00}\rangle_{3,4}=\frac{1}{\sqrt{d}}\sum_{j=0}^{d-1}\vert j\rangle_3\vert j\rangle_4$ and $\vert\psi_{00}\rangle_{5,6}=\frac{1}{\sqrt{d}}\sum_{j=0}^{d-1}\vert j\rangle_5\vert j\rangle_6$. Here, $P_1$ has particles 3 and 5, $P_2$ has particles 4 and 6.

$P_1$ ($P_2$) performs Bell measurement on particles 1 and 3 (2 and 6), and obtains the measurement $M_1=\{m_{11},m_{12}\}$ ($M_2=\{m_{21},m_{22}\}$), $m_{11},m_{12},m_{21},m_{22}\in{1,2,\ldots,d-1}$. At the same time, the particle 4 (5) owned by $P_2$ ($P_1$) collapses to the state $\vert\phi^\prime\rangle_4=U^{\dagger}(m_{11},m_{12})\vert\phi\rangle_4$ ($\vert\varphi^\prime\rangle_5=U^{\dagger}(m_{21},m_{22})\vert\phi\rangle_5$).

\textbf{Step 3: The encryption process of the transmitted particles.} $P_1$ ($P_2$) performs the unitary operation $U^{\dagger}(m_{11},m_{12})$ ($U^{\dagger}(m_{21},m_{22})$) on the particle $\vert\varphi^{\prime\prime}\rangle_5$ ($\vert\phi^\prime\rangle_4$), which is evolved to $\vert\varphi^{\prime\prime}\rangle_5=U^{\dagger}(m_{11},m_{12})U^{\dagger}(m_{21},m_{22})
\vert\phi\rangle_5$ ($\vert\phi^{\prime\prime}\rangle_4=U^{\dagger}(m_{21},m_{22})U^{\dagger}(m_{11},m_{12})
\vert\phi\rangle_4$). Then, $P_1$ ($P_2$) sends the particle $\vert\varphi^{\prime\prime}\rangle_5$ ($\vert\phi^{\prime\prime}\rangle_4$) to $Q_2$ ($Q_1$).

\textbf{Step 4: The homomorphic encryption process of the prepared particles (QFHE.Enc).}  $P_1$ and $P_2$ prepare the quantum states $\{\vert m_{11}\rangle_7,\vert m_{12}\rangle_8\}$ and $\{\vert m_{21}\rangle_9,\vert m_{22}\rangle_{10}\}$ in order based on their measurements $M_1$ and $M_2$. Then, according to the encryption keys $s_1$ and $s_2$, $P_1$ encodes particles 7 and 8. The encoded quantum states are $U(s_1,s_2)\vert m_{11}\rangle_7$ and $U(s_1,s_2)\vert m_{12}\rangle_8$, which are sent to $V_1$ by $P_1$. Similar to the operation performed by $P_1$, $P_2$ ciphers particles 9 and 10 with $s_1$ and $s_2$. The ciphered quantum states evolve into $U(s_1,s_2)\vert m_{21}\rangle_9$ and $U(s_1,s_2)\vert m_{22}\rangle_{10}$, they are sent to $V_1$ by $P_2$.

\textbf{Step 5: The homomorphic evaluation process of $V_1$ (QFHE.Eval).} After receiving the particles $U(s_1,s_2)\vert m_{11}\rangle_7$,  $U(s_1,s_2)\vert m_{12}\rangle_8$, $U(s_1,s_2)\vert m_{21}\rangle_9$ and $XU(s_1,s_2)\vert m_{22}\rangle_{10}$ sent by $P_1$ and $P_2$, $V_1$ performs the evaluation operation $CX_{7,9}$ on particles 7 and 9, while performing the evaluation operation $CX_{8,10}$ on particles 8 and 10. Subsequently, $V_1$ sends the evaluated particles 9 and 10 to $V_2$.

\textbf{Step 6: The homomorphic evaluation process of $V_2$ (QFHE.Eval).} Upon receiving particles 9 and 10 from $V_1$, $V_2$ prepares two particles 11 and 12 in the $\vert 0\rangle$ state. He then performs the operation $CX_{9,11}$ on particles 9 and 11, whilst performing the operation $CX_{10,12}$ on particles 10 and 12. Afterwards, $V_2$ sends the evaluated particles 9 and 10 to $Q_1$, as well as particles 11 and 12 to $Q_2$.

\textbf{Step 7: The homomorphic decryption process of the prepared particles (QFHE.Dec)} Depending on the above steps and the keys $s_1$ and $s_2$ shared among $P_1$, $P_2$, $Q_1$ and $Q_2$, $Q_1$ and $Q_2$ calculate their own decryption keys, which are $\{s_1, 2s_2\}$ and $\{0,2s_2\}$ respectively. In accordance with the decryption key $\{s_1, 2s_2\}$, $Q_1$ performs the decryption operation $U^\dagger(s_1,2s_2)$ on particles 9 and 10. The obtained quantum states are $\vert m_{11}+m_{21}\rangle_{9}$ and $\vert m_{12}+m_{22}\rangle_{10}$. Similarly, $Q_2$ performs the decryption operation $U^\dagger(0,2s_2)$ on particles 11 and 12 based on the decryption keys $\{0,2s_2\}$, getting the quantum states $\vert m_{11}+m_{21}\rangle_{11}$ and $\vert m_{12}+m_{22}\rangle_{12}$ respectively.

\textbf{Step 8: The decryption process of the transmitted particles.} $Q_1$ measures his particles 11 and 12 in the computational basis $B=\{\vert 0\rangle, \vert 1\rangle,\ldots,\vert d-1\rangle\}$, gaining sequentially measurements $m_{11}+m_{21}$ and $m_{12}+m_{22}$. Analogously, $Q_2$ measures particles 9 and 10 in the computational basis $B$, yields successively measurement results as $m_{11}+m_{21}$ and $m_{12}+m_{22}$. Subsequently, $Q_1$ and $Q_2$ perform respectively decryption operation $U(m_{11}+m_{21},m_{12}+m_{22})$ (i.e., $U(M_1+M_2)$) on the received particles $\vert\phi^{\prime\prime}\rangle_4$ and $\vert\varphi^{\prime\prime}\rangle_5$. Without considering the global phase, the quantum states acquired by $Q_1$ and $Q_2$ are $\vert\phi^{\prime\prime\prime}\rangle_4=\sum_{j=0}^{d-1}\alpha_{j}\vert j\rangle_4$ and $\vert\varphi^{\prime\prime\prime}\rangle_5=\sum_{j=0}^{d-1}\beta_{j}\vert j\rangle_5$ in order. That is, $Q_1$ and $Q_2$ recover the quantum states $\vert\phi\rangle_1$ and $\vert\varphi\rangle_2$ transmitted by $P_1$ and $P_2$, respectively.

\begin{figure*}[bp]
\centering
\begin{sideways}
\begin{minipage}{\textheight}
\centering
\includegraphics[width=\textwidth]{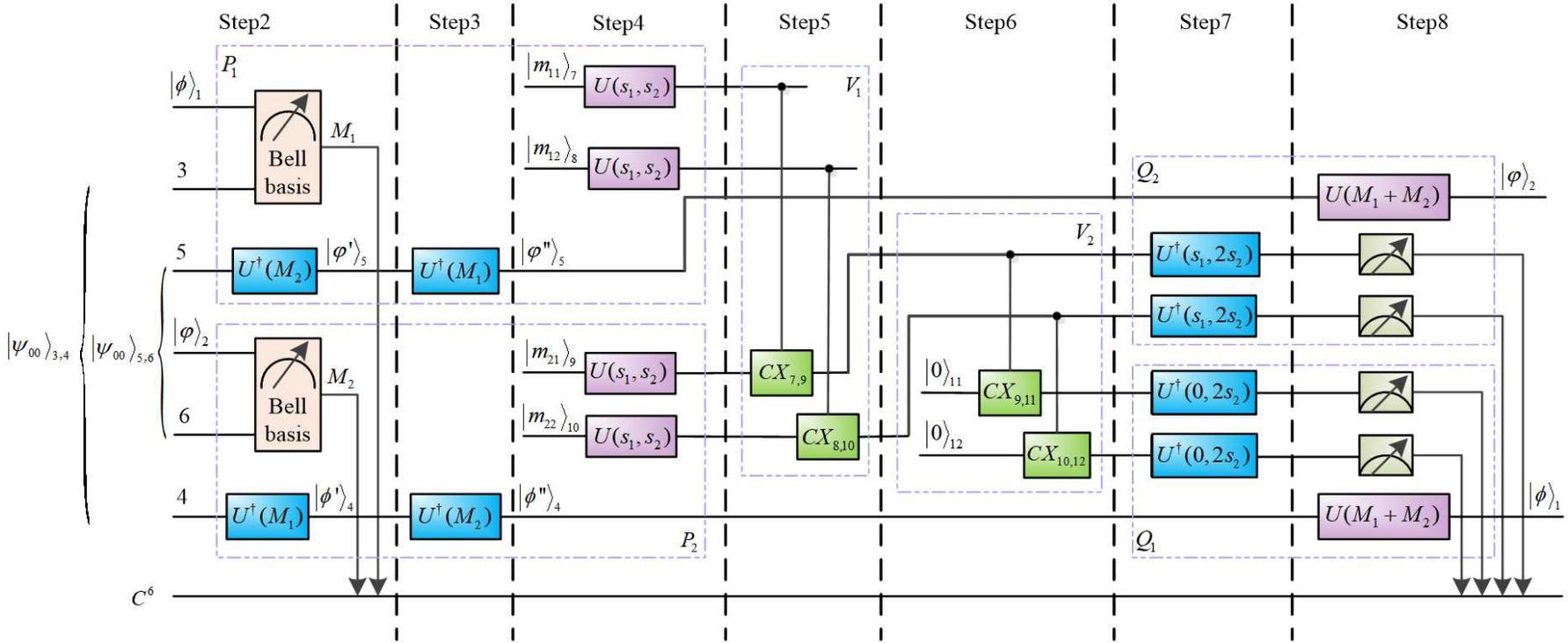}
\captionsetup{justification=centering,margin={0cm,0cm}}
\caption{The quantum circuit of the proposed quantum network coding protocol}\label{fig:5}
\end{minipage}
\end{sideways}
\end{figure*}

Through the above step 1 to step 8, $P_1$ and $P_2$ successfully transmit their respective quantum states $\vert\phi\rangle_1$ and $\vert\varphi\rangle_2$ to $Q_1$ and $Q_2$. In order to better understand the proposed protocol, we design its quantum circuit, as shown in Figure \ref{fig:5}.

Simultaneously, an example of dimension 3 (i.e. $d=3$) is given to further illustrate the execution of the protocol. For simplicity, the global phases in the example are ignored. In this example, two source nodes $P_1$ and P2 and two sink nodes Q1 and Q2 share keys $s_1$ and $s_2$. $P_1$ and $P_2$ have quantum states $\vert\phi\rangle_1=\sum_{j=0}^{2}\alpha_{j}\vert j\rangle_1$ and $\vert\varphi\rangle_2=\sum_{j=0}^{2}\beta_{j}\vert j\rangle_2$, respectively, and they share two pairs of Bell states $\vert\psi_{00}\rangle_{3,4}=\frac{1}{\sqrt{3}}\sum_{j=0}^{2}\vert j\rangle_3\vert j\rangle_4$ and $\vert\psi_{00}\rangle_{5,6}=\frac{1}{\sqrt{3}}\sum_{j=0}^{2}\vert j\rangle_5\vert j\rangle_6$. In step 2, if $P_1$ ($P_2$) gets Bell measurement result $M_1=\{0,1\}$ ($M_2=\{1,2\}$) for particles 1 and 3 (2 and 6), then the particle 4 (5) of $P_2$ ($P_1$) collapses to $\vert\phi'\rangle_4=\sum_{j=0}^{2}\alpha_{j}\vert j+1\rangle_4$ ( $\vert\varphi'\rangle_5=\sum_{j=0}^{2}\beta_{j}\vert j+2\rangle_5$). In step 3, $P_1$ and $P_2$ send the encrypted particles $\vert\varphi''\rangle_5=\sum_{j=0}^{2}\beta_{j}\omega^{-j}\vert j\rangle_5$ and $\vert\phi''\rangle_4=\sum_{j=0}^{2}\alpha_{j}\omega^{-j}\vert j\rangle_4$ to $Q_2$ and $Q_1$, respectively.

In step 4, according to the measurement results $M_1=\{0,1\}$ ($M_2=\{1,2\}$), $P_1$ ($P_2$) prepares the quantum states $\{\vert0\rangle_7,\vert1\rangle_8\}$ ($\{\vert1\rangle_9,\vert2\rangle_{10}\}$). Then, $P_1$ ($P_2$) encodes them by by making use of the keys $s_1=2$ and $s_2=1$, and obtains the encrypted particles $\{\omega\vert2\rangle_7,\vert0\rangle_8\}$ ($\{\vert0\rangle_9,\omega^{2}\vert1\rangle_{10}\}$). The particles 7, 8, 9 and 10 are sent by $P_1$ and $P_2$ to $V_1$. In step 5, $V_1$ carries out the evaluation operation $CX_{7,9}$ ($CX_{8,10}$) on the particles 7 and 9 (8 and 10). Afterwards, he sends the evaluated particles $\{\vert2\rangle_9,\omega^2\vert1\rangle_{10}\}$ to $V_2$. In step 6, after performing the evaluation operation $CX_{9,11}$ ($CX_{10,12}$) on particles 9 and 11 (10 and 12), $V_2$ sends the resulting evaluated particles $\{\vert2\rangle_9,\omega^2\vert1\rangle_{10}\}$ and $\{\vert2\rangle_{11},\vert1\rangle_{12}\}$ to $Q_2$ and $Q_1$, respectively. In step 7, according to the obtained decryption key $\{s_1=2, 2s_2=2\}$ ($\{0,2s_2=2\}$), $Q_1$ ($Q_2$) executes the decryption operation $U^\dagger(2,2)$ ($U^\dagger(0,2)$) on particles 9 and 10 (11 and 12), resulting in the quantum states $\{\vert1\rangle_{9},\vert0\rangle_{10}\}$ ($\{\vert1\rangle_{11},\vert0\rangle_{12}\}$).

At the end of the protocol, $Q_1$ and $Q_2$  utilize the computational basis $B=\{\vert0\rangle,\vert1\rangle,\vert2\rangle\}$ to measure the quantum states $\{\vert1\rangle_{9},\vert0\rangle_{10}\}$ and $\{\vert1\rangle_{11},\vert0\rangle_{12}\}$ in their hands, respectively. They both get the measurement result $\{1,0\}$. After that, depending on the measurement $\{1,0\}$, $Q_1$ and $Q_2$ execute respectively the decryption operation $U(1,0)$ on the received particles $\vert\phi''\rangle_4$ and $\vert\varphi''\rangle_5$. They get the quantum states $\vert\phi'''\rangle_4=\sum_{j=0}^{2}\alpha_{j}\vert j\rangle_4$ and $\vert\varphi'''\rangle_5=\sum_{j=0}^{2}\beta_{j}\vert j\rangle_5$ in turn. Obviously, $Q_1$ and $Q_2$ recover the quantum states $\vert\phi\rangle_1$ and $\vert\varphi\rangle_2$ transmitted by $P_1$ and $P_2$. Thus, $P_1$ and $P_2$ perfectly transmit their respective quantum states $\vert\phi\rangle_1$ and $\vert\varphi\rangle_2$ to $Q_1$ and $Q_2$.

\section{Protocol Analysis}\label{sec4}
In this section, the correctness of the proposed QNC protocol are firstly analyzed. Then, we analyze the security of the proposed QNC protocol.

\subsection{Correctness}\label{subsec4.1}

As can be seen from the example in Section 3.2, the source nodes $P_1$ and $P_2$ successfully transmit their respective quantum states to the corresponding sink nodes $Q_1$ and $Q_2$ in the ideal case. A rigorous proof of the correctness of the protocol from the theoretical point of view will be given as follows.

First of all, $P_1$ ($P_2$) performs Bell measurement on particles 1 and 3 (2 and 6), and the measurements result is ${{M}_{1}}=\{{{m}_{11}},{{m}_{12}}\}$ (${{M}_{2}}=\{{{m}_{21}},{{m}_{22}}\}$). At the same time, particle 5 (4) of $P_1$ ($P_2$) collapses to ${{U}^{\dagger}}({{m}_{21}},{{m}_{22}}){{\left| \varphi  \right\rangle }_{5}}$ (${{U}^{\dagger}}({{m}_{11}},{{m}_{12}}){{\left| \phi  \right\rangle }_{4}}$), which is encrypted by making use of the operation ${{U}^{\dagger}}({{m}_{11}},{{m}_{12}})$ (${{U}^{\dagger}}({{m}_{21}},{{m}_{22}})$). The final states of particles 4 and 5 are in

\begin{equation}
\begin{aligned}
{{\left| \phi^{\prime\prime} \right\rangle }_{4}}
 & ={{U}^{\dagger}}({{m}_{21}},{{m}_{22}}){{U}^{\dagger}}
({{m}_{11}},{{m}_{12}}){{\left| \phi  \right\rangle }_{4}} \\
 & =\sum\limits_{j=0}^{d-1}{{{\omega }^{-j{{m}_{21}}}}}\left| j+{{m}_{22}} \right\rangle \left\langle  j \right|\sum\limits_{k=0}^{d-1}{{{\omega }^{-k{{m}_{11}}}}}\left| k+{{m}_{12}} \right\rangle \left\langle  k \right|\sum\limits_{l=0}^{d-1}{{{\alpha }_{l}}{{\left| l \right\rangle }_{4}}} \\
 & =\sum\limits_{k=0}^{d-1}{{{\omega }^{-(k+{{m}_{12}}){{m}_{21}}}}{{\alpha }_{k}}{{\omega }^{-k{{m}_{11}}}}}{{\left| k+{{m}_{12}}+{{m}_{22}} \right\rangle }_{4}} \\
 & ={{\omega }^{-{{m}_{12}}{{m}_{21}}}}\sum\limits_{k=0}^{d-1}{{{\alpha }_{k}}{{\omega }^{-k({{m}_{11}}+{{m}_{21}})}}}{{\left| k+{{m}_{12}}+{{m}_{22}} \right\rangle }_{4}}
\end{aligned}
\label{eq:17}
\end{equation}
and
\begin{equation}
\begin{aligned}
 {{\left| \varphi^{\prime\prime} \right\rangle }_{5}}
 & ={{U}^{\dagger}}({{m}_{11}},{{m}_{12}}){{U}^{\dagger}}
 ({{m}_{21}},{{m}_{22}}){{\left| \varphi  \right\rangle }_{5}} \\
 & =\sum\limits_{j=0}^{d-1}{{{\omega }^{-j{{m}_{11}}}}}\left| j+{{m}_{12}} \right\rangle \left\langle  j \right|\sum\limits_{k=0}^{d-1}{{{\omega }^{-k{{m}_{21}}}}}\left| k+{{m}_{22}} \right\rangle \left\langle  k \right|\sum\limits_{l=0}^{d-1}{{{\beta }_{l}}{{\left| l \right\rangle }_{5}}} \\
 & =\sum\limits_{k=0}^{d-1}{{{\omega }^{-(k+{{m}_{22}}){{m}_{11}}}}{{\beta }_{k}}{{\omega }^{-k{{m}_{21}}}}}{{\left| k+{{m}_{12}}+{{m}_{22}} \right\rangle }_{5}} \\
 & ={{\omega }^{-{{m}_{11}}{{m}_{22}}}}\sum\limits_{k=0}^{d-1}{{{\beta }_{k}}{{\omega }^{-k({{m}_{11}}+{{m}_{21}})}}}{{\left| k+{{m}_{12}}+{{m}_{22}} \right\rangle }_{5}}.
\end{aligned}
\label{eq:18}
\end{equation}

Second, $P_1$ ($P_2$) encodes the prepared quantum state $\{\vert m_{11}\rangle_{7},\vert m_{12}\rangle_{8}\}$ ($\{\vert m_{21}\rangle_{9},\vert m_{22}\rangle_{10}\}$) with the keys $s_1$ and $s_2$, and obtains the quantum states $\{U({{s}_{1}},{{s}_{2}}){{\left| {{m}_{11}} \right\rangle }_{7}},U({{s}_{1}},{{s}_{2}}){{\left| {{m}_{12}} \right\rangle }_{8}}\}$ ($\{U({{s}_{1}},{{s}_{2}}){{\left| {{m}_{21}} \right\rangle }_{9}}, U({{s}_{1}},{{s}_{2}}){{\left| {{m}_{22}} \right\rangle }_{10}}$\}). When $V_1$ receives these four particles, he performs the operation $CX_{7,9}$ ($CX_{8,10}$) on particles 7 and 9 (8 and 10). According to Eq. (\ref{eq:10}), the states of the system are in

\begin{equation}
\begin{aligned}
& C{{X}_{7,9}}\left( U({{s}_{1}},{{s}_{2}}){{\left| {{m}_{11}} \right\rangle }_{7}}\otimes U({{s}_{1}},{{s}_{2}}){{\left| {{m}_{21}} \right\rangle }_{9}} \right) \\
 & =\left( U(2{{s}_{1}},{{s}_{2}})\otimes U({{s}_{1}},2{{s}_{2}}) \right)C{{X}_{7,9}}\left( {{\left| {{m}_{11}} \right\rangle }_{7}}\otimes {{\left| {{m}_{21}} \right\rangle }_{9}} \right) \\
 & =U(2{{s}_{1}},{{s}_{2}}){{\left| {{m}_{11}} \right\rangle }_{7}}\otimes U({{s}_{1}},2{{s}_{2}}){{\left| {{m}_{11}}+{{m}_{21}} \right\rangle }_{9}}, \\
 & C{{X}_{8,10}}\left( U({{s}_{1}},{{s}_{2}}){{\left| {{m}_{12}} \right\rangle }_{8}}\otimes U({{s}_{1}},{{s}_{2}}){{\left| {{m}_{22}} \right\rangle }_{10}} \right) \\
 & =C{{X}_{8,10}}\left( U({{s}_{1}},{{s}_{2}})\otimes U({{s}_{1}},{{s}_{2}}) \right)\left( {{\left| {{m}_{12}} \right\rangle }_{8}}\otimes {{\left| {{m}_{22}} \right\rangle }_{10}} \right) \\
 & =U(2{{s}_{1}},{{s}_{2}}){{\left| {{m}_{12}} \right\rangle }_{8}}\otimes U({{s}_{1}},2{{s}_{2}}){{\left| {{m}_{12}}+{{m}_{22}} \right\rangle }_{10}}.
\end{aligned}
\label{eq:19}
\end{equation}
Upon $V_2$ performing operation $CX_{9,11}$ ($CX_{10,12}$) on the received particle 9 (10) and the prepared particle 11 (12), similarly according to Eq. (\ref{eq:10}), the states of the system evolve respectively as

\begin{equation}
\begin{aligned}
 & C{{X}_{9,11}}\left( U({{s}_{1}},2{{s}_{2}}){{\left| {{m}_{11}}+{{m}_{21}} \right\rangle }_{9}}\otimes {{\left| 0 \right\rangle }_{11}} \right) \\
 & =\left( U({{s}_{1}},2{{s}_{2}})\otimes U(0,2{{s}_{2}}) \right)C{{X}_{9,11}}({{\left| {{m}_{11}}+{{m}_{21}} \right\rangle }_{9}}\otimes {{\left| 0 \right\rangle }_{11}}) \\
 & =U({{s}_{1}},2{{s}_{2}}){{\left| {{m}_{11}}+{{m}_{21}} \right\rangle }_{9}}\otimes U(0,2{{s}_{2}}){{\left| {{m}_{11}}+{{m}_{21}} \right\rangle }_{11}}, \\
 & C{{X}_{10,12}}\left( U({{s}_{1}},2{{s}_{2}}){{\left| {{m}_{12}}+{{m}_{22}} \right\rangle }_{10}}\otimes {{\left| 0 \right\rangle }_{12}} \right) \\
 & =\left( U({{s}_{1}},2{{s}_{2}})\otimes U(0,2{{s}_{2}}) \right)C{{X}_{10,12}}\left( {{\left| {{m}_{12}}+{{m}_{22}} \right\rangle }_{10}}\otimes {{\left| 0 \right\rangle }_{12}} \right) \\
 & =U({{s}_{1}},2{{s}_{2}}){{\left| {{m}_{12}}+{{m}_{22}} \right\rangle }_{10}}\otimes U(0,2{{s}_{2}}){{\left| {{m}_{12}}+{{m}_{22}} \right\rangle }_{12}}. \\
\end{aligned}
\label{eq:20}
\end{equation}

At last, using the decryption keys $\{s_1, 2s_2\}$ ($\{0,2s_2\}$), $Q_1$ ($Q_2$) performs the decryption operation $U^\dagger(s_1,2s_2)$ ($U^\dagger(0,2s_2)$) on the received particles 9 and 10 (11 and 12). The yielded quantum states are $\left\{ {{\left| {{m}_{11}}+{{m}_{21}} \right\rangle }_{9}},{{\left| {{m}_{11}}+{{m}_{21}} \right\rangle }_{10}} \right\}$ ($\left\{ {{\left| {{m}_{11}}+{{m}_{21}} \right\rangle }_{11}},{{\left| {{m}_{12}}+{{m}_{22}} \right\rangle }_{12}} \right\}$). Based on the measurement result $\left\{ {{m}_{11}}+{{m}_{21}},{{m}_{12}}+{{m}_{22}} \right\}$, $Q_1$ and $Q_2$ execute respectively the operation $U({{m}_{11}}+{{m}_{21}},{{m}_{12}}+{{m}_{22}})$ on the particles ${{\left| \phi^{\prime\prime} \right\rangle }_{4}}$ and ${{\left| \varphi^{\prime\prime} \right\rangle }_{5}}$. The eventually obtained quantum states are in order
\begin{equation}
\begin{aligned}
{{\left| \phi^{\prime\prime\prime} \right\rangle }_{4}}  &=U({{m}_{11}}+{{m}_{21}},{{m}_{12}}+{{m}_{22}}){{\left| {{\phi }_{1}} \right\rangle }_{4}} \\
 & =\sum\limits_{j=0}^{d-1}{{{\omega }^{j({{m}_{11}}+{{m}_{21}})}}}\left| j \right\rangle \left\langle  j+{{m}_{12}}+{{m}_{22}} \right|{{\omega }^{-{{m}_{12}}{{m}_{21}}}} \\
 &\sum\limits_{k=0}^{d-1}{{{\alpha }_{k}}{{\omega }^{-k({{m}_{11}}+{{m}_{21}})}}}{{\left| k+{{m}_{12}}+{{m}_{22}} \right\rangle }_{4}} \\
 & ={{\omega }^{-{{m}_{12}}{{m}_{21}}}}\sum\limits_{j=0}^{d-1}{{{\alpha }_{j}}{{\left| j \right\rangle }_{4}}}
\end{aligned}
\label{eq:21}
\end{equation}
and
\begin{equation}
\begin{aligned}
 {{\left| \varphi^{\prime\prime\prime} \right\rangle }_{5}}
 &=U({{m}_{11}}+{{m}_{21}},{{m}_{12}}+{{m}_{22}}){{\left| {{\varphi }_{1}} \right\rangle }_{5}} \\
 & =\sum\limits_{j=0}^{d-1}{{{\omega }^{j({{m}_{11}}+{{m}_{21}})}}}\left| j \right\rangle \left\langle  j+{{m}_{12}}+{{m}_{22}} \right|{{\omega }^{-{{m}_{11}}{{m}_{22}}}} \\
 &\sum\limits_{k=0}^{d-1}{{{\beta }_{k}}{{\omega }^{-k({{m}_{11}}+{{m}_{21}})}}}{{\left| k+{{m}_{12}}+{{m}_{22}} \right\rangle }_{5}} \\
 & ={{\omega }^{-{{m}_{11}}{{m}_{22}}}}\sum\limits_{j=0}^{d-1}{{{\beta }_{j}}}{{\left| j \right\rangle }_{5}}.
\end{aligned}
\label{eq:22}
\end{equation}

From Eq. (\ref{eq:21}) and (\ref{eq:22}), it can be seen that $Q_1$ and $Q_2$ recover the quantum states ${{\left| \phi  \right\rangle }_{1}}$ and ${{\left| \varphi  \right\rangle }_{2}}$ transmitted by $P_1$ and $P_2$, without considering the global phase. As a summary, the proposed QNC protocol is correct.

\subsection{Security}\label{subsec4.2}

In this section, we first give the security definition that the fully mixed state guarantees the security of the QNC scheme \cite{bibLM,bibZS}. Then, on that basis, the security of the proposed QNC protocol is proved. That is, it is resistant to attacks launched from both internal dishonest intermediate nodes and external attackers.

\textbf{Definition 1} ({\em Information-theoretic security}) If for all input states $\sigma$, its output state $\sigma_{c}$ is completely mixed, then the algorithm is secure. The relationship between the input state $\sigma$ and the output state $\sigma_{c}$ is described as follows:

\begin{equation}
\begin{aligned}
{{\sigma }_{c}}=\frac{1}{{{d}^{2}}}\sum\limits_{{{s}_{1}},{{s}_{2}}\in \{0,1,\cdots ,d-1\}}{U({{s}_{1}},{{s}_{2}})\sigma {{U}^{\dagger }}({{s}_{1}},{{s}_{2}})}=\frac{{{I}_{d}}}{d}.
\end{aligned}
\label{eq:23}
\end{equation}
Here, $\sigma$ denotes the density matrix of all possible input states, $U({{s}_{1}},{{s}_{2}})$ (${{s}_{1}},{{s}_{2}}\in \{0,1,2,\ldots ,d-1\}$) is the unitary transformation between all input states and output states.

In the proposed QNC protocol, if an attacker wants to recover the quantum states sent by the two source nodes $P_1$ and $P_2$ to the two sink nodes $Q_1$ and $Q_2$ respectively, he needs to obtain the measurements ${{M}_{1}}=\{{{m}_{11}},{{m}_{12}}\}$, ${{M}_{2}}=\{{{m}_{21}},{{m}_{22}}\}$ or ${{M}_{1}}+{{M}_{2}}$, with ${{m}_{11}},{{m}_{12}},{{m}_{21}},{{m}_{22}}\in \mathbb{Z}_{d}$. The attacker can acquire $M_1$, $M_2$ or $M_1+M_2$ by the following three ways.

(T1) Intercepting the quantum states sent by $P_1$ and $P_2$ to V1

According to the Definition 1, it is clear that this phase is secure. Its proof is as follows. $\sigma =\sum\limits_{l=0}^{d-1}{{{a}_{l}}\left| l \right\rangle \left\langle  l \right|}$ is the density matrix of all possible plaintexts of the quantum states $\left\{ {{\left| {{m}_{11}} \right\rangle }_{7}},\left| {{m}_{12}} \right\rangle {}_{8},{{\left| {{m}_{21}} \right\rangle }_{9}},{{\left| {{m}_{22}} \right\rangle }_{10}} \right\}$ prepared by $P_1$ and $P_2$ in step 4. Assume that the attacker cannot know the probability distribution associated with the transmitted quantum state in advance through other channels, thus the quantum state plaintexts in front of the attacker have the same probability ${{a}_{l}}=1/d$, $l\in \{0,1,\ldots ,d-1\}$. $s_1$ and $s_2$ are the keys shared by $P_1$, $P_2$, $Q_1$ and $Q_2$. The keys are only held by these four parties and are not available to the attacker. When the attacker tries to intercept the quantum states sent by $P_1$ and $P_2$ to $V_1$, because the output

\begin{equation}
\begin{aligned}
  {{\sigma }_{c}}&=\frac{1}{{{d}^{2}}}\sum\limits_{{{s}_{1}},{{s}_{2}}\in \{0,1,\cdots ,d-1\}}{U({{s}_{1}},{{s}_{2}})\sigma {{U}^{\dagger }}({{s}_{1}},{{s}_{2}})} \\
 & =\frac{1}{{{d}^{2}}}\sum\limits_{{{s}_{1}},{{s}_{2}}\in \{0,1,\cdots ,d-1\}}{\sum\limits_{j=0}^{d-1}{{{\omega }^{{{s}_{1}}j}}\left| j \right\rangle \left\langle  j+{{s}_{2}} \right|\sum\limits_{l=0}^{d-1}{{{a}_{l}}\left| l \right\rangle \left\langle  l \right|\sum\limits_{j=0}^{d-1}{{{\omega }^{-{{s}_{1}}j}}\left| j+{{s}_{2}} \right\rangle \left\langle  {{j}_{1}} \right|}}}} \\
 & =\frac{1}{{{d}^{2}}}\sum\limits_{{{s}_{1}},{{s}_{2}}\in \{0,1,\cdots ,d-1\}}{\sum\limits_{l=0}^{d-1}{{{a}_{l}}{{\omega }^{{{s}_{1}}(l-{{s}_{2}})}}\left| l-{{s}_{2}} \right\rangle \left\langle  l \right|\sum\limits_{j=0}^{d-1}{{{\omega }^{-{{s}_{1}}j}}\left| j+{{s}_{2}} \right\rangle \left\langle  j \right|}}} \\
 & =\frac{1}{{{d}^{2}}}\sum\limits_{{{s}_{1}},{{s}_{2}}\in \{0,1,\cdots ,d-1\}}{\sum\limits_{l=0}^{d-1}{{{a}_{l}}\left| l-{{s}_{2}} \right\rangle \left\langle  l-{{s}_{2}} \right|}} \\
 & =\frac{1}{{{d}^{2}}}\sum\limits_{{{s}_{1}},{{s}_{2}}\in \{0,1,\cdots ,d-1\}}{{{a}_{l}}{{I}_{d}}} \\
 & =\frac{{{I}_{d}}}{d}
\end{aligned}
\label{eq:24}
\end{equation}
of the encryption algorithm is a fully mixed state, therefore the attacker can only intercept or detect the fully mixed quantum states. As a result, the attacker cannot obtain any valued information about $\left\{ {{m}_{11}},{{m}_{12}},{{m}_{21}},{{m}_{22}} \right\}$ from the intercepted quantum states.

(T2) Intercepting the quantum states sent from $V_1$ to $V_2$

Since the quantum states $\{\left| {{m}_{11}} \right\rangle _{7},\left| {{m}_{12}} \right\rangle_{8},\left| {{m}_{21}} \right\rangle_{9},\left| {{m}_{22}} \right\rangle_{10}\} $ are prepared based on Bell measurement results, these quantum states can be expressed uniformly as $\left| {{m}_{a,b}} \right\rangle =\frac{1}{\sqrt{d}}\sum\limits_{j=0}^{d-1}{\left| j \right\rangle }$ ($a,b\in \{1,2\}$) in combination with the analysis in (T1). As can be seen from Eq. (\ref{eq:19}), the operation of particles 7 and 9 is the same as that of 8 and 10. Hence, we take particles 7 and 9 as an example for our analysis. After the homomorphic evaluation, the density operator of the system in which particles 7 and 9 are located is

\begin{equation}
\begin{aligned}
{{\rho }_{7,9}}&=((U(2{{s}_{1}},{{s}_{2}}){{\left| {{m}_{11}} \right\rangle }_{7}}\otimes U({{s}_{1}},2{{s}_{2}}){{\left| {{m}_{11}}+{{m}_{21}} \right\rangle }_{9}}) \\
 & {{\left( U(2{{s}_{1}},{{s}_{2}}){{\left| {{m}_{11}} \right\rangle }_{7}}\otimes U({{s}_{1}},2{{s}_{2}}){{\left| {{m}_{11}}+{{m}_{21}} \right\rangle }_{9}} \right)}^{\dagger }} ) \\
 & =U(2{{s}_{1}},{{s}_{2}}){{\left| {{m}_{11}} \right\rangle }_{7}}_{7}\left\langle  {{m}_{11}} \right|{{U}^{\dagger }}(2{{s}_{1}},{{s}_{2}})\otimes U({{s}_{1}},2{{s}_{2}}) \\
 &{{\left| {{m}_{11}}+{{m}_{21}} \right\rangle }_{9}}
  _{9}\left\langle  {{m}_{11}}+{{m}_{21}} \right|{{U}^{\dagger }}({{s}_{1}},2{{s}_{2}}).
\end{aligned}
\label{eq:25}
\end{equation}
Tracing out the particle 7, we can get the reduced density operator of the particle 9, which is

\begin{equation}
\begin{aligned}
 {{\rho }_{9}}&=t{{r}_{7}}\left( {{\rho }_{7,9}} \right) \\
 & =U({{s}_{1}},2{{s}_{2}}){{\left| {{m}_{11}}+{{m}_{21}} \right\rangle }_{9}}_{9}\left\langle  {{m}_{11}}+{{m}_{21}} \right|{{U}^{\dagger }}({{s}_{1}},2{{s}_{2}}) \\
 & t{{r}_{7}}\left( U(2{{s}_{1}},{{s}_{2}}){{\left| {{m}_{11}} \right\rangle }_{7}}_{7}\left\langle  {{m}_{11}} \right|{{U}^{\dagger }}(2{{s}_{1}},{{s}_{2}}) \right) \\
 & =\frac{1}{d}U({{s}_{1}},2{{s}_{2}})\sum\limits_{j=0}^{d-1}{{{\left| i+j \right\rangle }_{9}}_{9}\left\langle  i+j \right|}{{U}^{\dagger }}({{s}_{1}},2{{s}_{2}}) \\
 & \frac{1}{d}\sum\limits_{i=0}^{d-1}{_{7}\left\langle  i \right|{{U}^{\dagger }}(2{{s}_{1}},{{s}_{2}})U(2{{s}_{1}},{{s}_{2}})\sum\limits_{i=0}^{d-1}{{{\left| i \right\rangle }_{7}}}} \\
 & =\frac{1}{d}{{I}_{d}}.
\end{aligned}
\label{eq:26}
\end{equation}

In the same way, we can obtain the reduced density operator for particle 10, that is
\begin{equation}
\begin{aligned}
 {{\rho }_{10}}&=t{{r}_{8}}\left( \left( U(2{{s}_{1}},{{s}_{2}}){{\left| {{m}_{12}} \right\rangle }_{8}}\otimes U({{s}_{1}},2{{s}_{2}}){{\left| {{m}_{12}}+{{m}_{22}} \right\rangle }_{10}} \right) \right. \\
 & \left. {{\left( U(2{{s}_{1}},{{s}_{2}}){{\left| {{m}_{12}} \right\rangle }_{8}}\otimes U({{s}_{1}},2{{s}_{2}}){{\left| {{m}_{12}}+{{m}_{22}} \right\rangle }_{10}} \right)}^{\dagger }} \right) \\
 & =\frac{1}{d}{{I}_{d}}.
\end{aligned}
\label{eq:27}
\end{equation}
When the attacker intercepts particles 9 and 10, since both the reduced density operators $\rho_9$ and $\rho_{10}$ are in the maximum mixed state, he cannot obtain any useful information about $m_{11}+m_{21}$ and $m_{12}+m_{22}$ from them.

(T3) Intercepting the quantum states sent to $Q_1$ and $Q_2$ by $V_2$

Following Eq. (\ref{eq:20}) and the analytical method of (T2), we find the reduced density operators of the particles 11 and 12,

\begin{equation}
\begin{aligned}
 {{\rho }_{11}}&=t{{r}_{9}}\left( \left( U({{s}_{1}},2{{s}_{2}}){{\left| {{m}_{11}}+{{m}_{21}} \right\rangle }_{9}}\otimes U(0,2{{s}_{2}}){{\left| {{m}_{11}}+{{m}_{21}} \right\rangle }_{11}} \right. \right) \\
 & \left. {{\left( U({{s}_{1}},2{{s}_{2}}){{\left| {{m}_{11}}+{{m}_{21}} \right\rangle }_{9}}\otimes U(0,2{{s}_{2}}){{\left| {{m}_{11}}+{{m}_{21}} \right\rangle }_{11}} \right)}^{\dagger }} \right) \\
 & =\frac{1}{d}{{I}_{d}}, \\
 {{\rho }_{12}}&=t{{r}_{10}}\left( \left( U({{s}_{1}},2{{s}_{2}}){{\left| {{m}_{12}}+{{m}_{22}} \right\rangle }_{10}}\otimes U(0,2{{s}_{2}}){{\left| {{m}_{12}}+{{m}_{22}} \right\rangle }_{12}} \right) \right. \\
 & \left. {{\left( U({{s}_{1}},2{{s}_{2}}){{\left| {{m}_{12}}+{{m}_{22}} \right\rangle }_{10}}\otimes U(0,2{{s}_{2}}){{\left| {{m}_{12}}+{{m}_{22}} \right\rangle }_{12}} \right)}^{\dagger}} \right) \\
 & =\frac{1}{d}{{I}_{d}}.
\end{aligned}
\label{eq:28}
\end{equation}
While intercepting particles 9, 10, 11 and 12, the attacker cannot obtain effective information about $m_{11}+m_{21}$ and $m_{12}+m_{22}$ from them due to the fact that $\rho_9$, $\rho_{10}$, $\rho_{11}$ and $\rho_{12}$ are in the maximum mixed state.

As a conclusion, at any stage of the above security analysis, the attacker cannot obtain any valuable information about $M_1$, $M_2$ or $M_1+M_2$ since the transmitted quantum states are all in the maximum mixed state. It is important to note that even if the attackers are dishonest intermediate nodes, $V_1$ and $V_2$, they are unable to obtain any valuable information about $M_1$, $M_2$ or $M_1+M_2$ during the above stages, whether they attack alone or conspire to attack. Therefore, the proposed quantum network coding protocol can resist both external attacks and internal attacks launched by dishonest intermediate nodes.

\section{conclusions}\label{sec5}

Before deriving our conclusions, we briefly discuss some advantages of the proposed QNC protocol compared with Refs. \cite{bibSZ,bibSP,bibOK,bibCC,bibLS}. Firstly, we proposed the first $d$-dimensional QFHE protocol based on universal quantum gates. It can be applied to various quantum cryptographic protocols, such as quantum secure direct communication, quantum secure multi-party summation, quantum network coding, etc. Secondly, the internal attacks launched by dishonest intermediate nodes, which are inevitable in practical applications, are taken into account in our protocol compared to protocols \cite{bibSP,bibOK,bibLS}. To this end, we design a $d$-dimensional QFHE protocol, which is utilised by the two source nodes to encrypt the prepared quantum states. This makes our protocol resistant not only to external attacks, but also to internal attacks. Thirdly, the communication efficiency of our protocol is much higher. The protocol \cite{bibCC} employs a key of length 8 to cipher the quantum states, while the proposed protocol requires only a key of length 2. In addition, since $d$-dimensional quantum system have a higher capacity than 2-dimensional quantum system. Therefore, these significantly improve the communication efficiency of the proposed protocol. Thirdly, the proposed protocol has a lower communication complexity. Compared to the quantum gates required for the encryption and decryption phases in the protocol \cite{bibCC}, our protocol requires only 1/2 of the original quantum gates, which reduces the communication complexity of the overall protocol. Fourthly, the generality of the proposed protocol is higher. The protocols \cite{bibSZ,bibSP,bibCC,bibLS} utilize 2-dimensional quantum resources to transmit quantum states. Whereas the proposed protocol implements the transmission of quantum states on $d$-dimensional quantum systems. Therefore, our protocol has higher generality.

In summary, we propose an efficient and secure QNC protocol with QFHE. At first, we construct a QFHE protocol based on $d$-dimensional universal quantum gates, and prove that the scheme is correct from the theoretical perspective. Based on this, we propose an efficient quantum network coding protocol, which achieves perfect transmission of quantum states over the butterfly network by making use of $d$-dimensional Bell states. In this protocol, two source nodes encrypt their respective prepared quantum states utilizing the quantum homomorphic encryption scheme, and send them to the first intermediate node. Next, the first intermediate node performs the homomorphic evaluation operation on the received quantum states. He then sends the target quantum states to the second intermediate node. Afterwards, the second intermediate node performs the homomorphic evaluation operation on all the quantum states in hand, and sends the control quantum states and the target quantum states to the two sink nodes respectively. Finally, the two sink nodes recover the quantum states transmitted by each of the two source nodes based on their measurement results. The performance analysis of the protocol shows that the proposed protocol is correct and can resist external attacks as well as internal attacks launched by dishonest intermediate nodes.

\ack {This work was supported by National Natural Science Foundation of China (Grants No. 61772134, No. 61976053, and No. 62171131), Fujian Province Natural Science Foundation (Grant No. 2022J01186), and Program for New Century Excellent Talents in Fujian Province University.}

\end{document}